\documentstyle[11pt,epsfig]{article}
\title{\bf Non-commutative multi-dimensional cosmology}
\author{N. Khosravi$^1$\thanks{email:
n-khosravi@sbu.ac.ir}, S. Jalalzadeh$^1$\thanks{email:
s-jalalzadeh@sbu.ac.ir}
  and H. R. Sepangi$^{1,2}$\thanks{email:
hr-sepangi@sbu.ac.ir}
\\ $^1${\small Department of Physics, Shahid Beheshti University, Evin,
Tehran 19839, Iran}\\$^2${\small Institute for Studies in
Theoretical Physics and Mathematics, P.O. Box 19395-5746, Tehran,
Iran }}
\begin{document}
\maketitle
\begin{abstract}
A non-commutative multi-dimensional cosmological model is
introduced and used to address the issues of compactification and
stabilization of extra dimensions and the cosmological constant
problem. We show that in such a scenario these problems find
natural solutions in a universe described by an increasing time
parameter.\vspace{5mm}\noindent
\end{abstract}
\section{Introduction}
Non-commutativity between space-time coordinates, first introduced
in \cite{1}, has been attracting considerable attention in the
recent past \cite{2,3,4}. This renewed interest has its roots in
the development of string and M-theories, \cite{5,6}. However, in
all fairness, investigation of non-commutative theories may also
be justified in its own right because of the interesting
predictions regarding, for example, the IR/UV mixing and
non-locality \cite{7}, Lorentz violation \cite{8} and new physics
at very short distance scales \cite{9,10,11}. The impact of
non-commutativity in cosmology has also been considerable and has
been addressed in different works \cite{12}. Hopefully,
non-commutative cosmology would lead us to the formulation of
semiclassical approximations of quantum gravity and tackles the
cosmological constant problem \cite{13}. Of particular interest
would be the application of non-commutativity to multi-dimensional
cosmology.

Multi-dimensional spaces were introduced for the geometric
unification of interactions by Kaluza and Klein and have since
been  a source of inspiration for numerous authors \cite{14}.
Also, the introduction of extra dimensions suggests possible
solutions to some of the important problems that cosmology has
been facing, namely, the mass hierarchy and cosmological constant
problem, to name but a few. That said, it should also be mentioned
that the question of  compactification and the stabilization of
the extra dimensions is a challenge one cannot avoid. In this
paper, we have considered a multi-dimensional cosmology with a
cosmological constant and introduced non-commutativity between the
scale factors of our ordinary space and the extra dimensions. We
have shown that the classical cosmology of this model can be
solved exactly. These solutions offer an explanation for the
cosmological constant problem and show how the extra dimensions
can be compactified.
\section{The Model}
We consider a cosmological model in which the space-time is
assumed to be of FRW type with a $d$-dimensional internal space.
The corresponding metric can be written as
\begin{equation}\label{1}
ds^2=-dt^2+\frac{R^2(t)}{\left(1+\frac{k}{4}r^2\right)^2}
(dr^2+r^2 d\Omega^2)+a^2(t)g^{(d)}_{ij}dx^idx^j
    \hspace{.15cm},
\end{equation}
where the total number of dimensions is $D=4+d$, $k=1,0,-1$
represents the usual spatial curvature, $R(t)$ and $a(t)$ are the
scale factors of the universe and the radius of the
$d$-dimensional internal space respectively and $g^{(d)}_{ij}$ is
the metric associated with the internal space, assumed to be
Ricci-flat. The Ricci scalar corresponding to metric (\ref{1}) is
\begin{equation}\label{2}
{\cal
R}=6\left[\frac{\ddot{R}}{R}+\frac{k+\dot{R}^{2}}{R^2}\right]+
2\,d\frac{\ddot{a}}{a}+d(d-1)\left(\frac{\dot{a}}{a}\right)^2\
+6\,d\frac{\dot{a}\dot{R}}{aR} ,
\end{equation}
where a dot represents differentiation with respect to $t$. Let
$a_0$ be the compactification scale of the internal space at
present time and
\begin{equation}\label{3}
v_d\equiv v_0\times v_i\equiv
a_0^d\times\int_{M_d}d^dx\sqrt{-g^{(d)}},
\end{equation}
the corresponding total volume of the internal space. Substitution
of equation (\ref{2}) and use of definition (\ref{3}) in the
Einstein-Hilbert action functional with a $D$-dimensional
cosmological constant $\Lambda$
\begin{equation}\label{4}
{\cal{S}}=\frac{1}{2k_D^2}\int_M d^Dx \sqrt{-g}({\cal R}
-2{\Lambda})+{\cal{S}}_{YGH},
\end{equation}
where $k_D$ is the $D$-dimensional gravitational constant and
${\cal{S}}_{YGH}$ is the York-Gibbons-Hawking boundary term, leads
to, after dimensional reduction
\begin{equation}\label{5}
{\cal S}=-v_{D-1}\int dt \left\{6\dot{R}^2\Phi
R+6\dot{R}\dot{\Phi}R^2+\frac{d-1}{d}\frac{\dot{\Phi}^2}{\Phi}R^3-6k\Phi
R-2\Phi R^3\Lambda\right\},
\end{equation}
where
\begin{equation}\label{6}
\Phi=\left(\frac{a}{a_0}\right)^d,
\end{equation}
and we have set $v_{D-1}=1$. To make the Lagrangian manageable,
consider the following change of variables
\begin{equation}\label{8}
\Phi R^3=\Upsilon^2(x_1^2-x_2^2),
\end{equation}
where $R=R(x_1,x_2)$ and $\Phi=\Phi(x_1,x_2)$ are functions of new
variables $x_1$, $x_2$. Let
\begin{eqnarray}\label{9}
\left\{%
\begin{array}{lll}
\Phi^{\rho_+}R^{\sigma_-}=\Upsilon(x_1+x_2),\\
\\
\Phi^{\rho_-}R^{\sigma_+}=\Upsilon(x_1-x_2), \\
\end{array}%
\right.
\end{eqnarray}
such that for $d\neq3$, $\Upsilon=1$ and we have
\begin{eqnarray}\label{10}
\left\{%
\begin{array}{ll}
\rho_\pm=\frac{1}{2}\pm \frac{3}{4}\sqrt{\frac{d+2}{3d}}\mp
\frac{1}{4\sqrt{\frac{d+2}{3d}}},\\
\sigma_{\pm}=\frac{1}{2}\left(3\mp \frac{1}{\sqrt{\frac{d+2}{3d}}}\right),\\
\end{array}
\right.
\end{eqnarray}
while for $d=3$, $\Upsilon=\frac{3}{\sqrt{5}}$ and
\begin{eqnarray}\label{101}
\left\{%
\begin{array}{ll}
\rho_\pm=\frac{1}{2}\pm \frac{\sqrt{5}}{10},\\
\sigma_{\pm}=3\left(\frac{1}{2}\pm \frac{\sqrt{5}}{10}\right).\\
\end{array}
\right.
\end{eqnarray}
Using the above transformations introduced in \cite{shahram} and
concentrating on $k=0$, the Lagrangian becomes
\begin{equation}\label{11}
{\cal
L}=-4\left(\frac{d+2}{d+3}\right)\left\{\dot{x_1}^2-\dot{x_2}^2-
\frac{\Lambda}{2}\left(\frac{d+3}{d+2}\right)\left(x_1^2-x_2^2\right)\right\}.
\end{equation}
Up to an overall  constant coefficient, we can write the effective
Hamiltonian as
\begin{equation}\label{14}
{\cal
H}=\frac{p_1^2}{4}-\frac{p_2^2}{4}+\omega^2\left(x_1^2-x_2^2\right),
\end{equation}
where $\omega^2$ is
\begin{equation}\label{1a}
\omega^2=\frac{1}{2}\left(\frac{d+3}{d+2}\right)\Lambda.
\end{equation}

The role of the variables $x_1$ and $x_2$ can be grasped easily if
one multiplies equation (\ref{8}) by $\Lambda$. Equations
(\ref{14}) and (\ref{1a}) then show that the potential energy for
our system of harmonic oscillators is proportional to the vacuum
energy $\Lambda$ times the volume of the multidimensional
universe. Also, as has been discussed  in \cite{17}, the wrong
sign in the Hamiltonian for the $x_2$ component is no cause for
concern since the equations of motion resulting from the
Hamiltonian are similar to those describing a system of two
ordinary uncoupled harmonic oscillators. In other words, the
potential in equation (\ref{14}) is given by
$V=\omega^2(x_1^2-x_2^2)$ so that the equations of motion become
\begin{eqnarray}
\ddot{x_1}=-\frac{\partial V}{\partial x_1} \hspace{3mm}
\mbox{and}\hspace{3mm} \ddot{x_2}=\frac{\partial V}{\partial x_2}.
\end{eqnarray}
For the $x_1$ component, the force is given by minus the gradient
of the potential, whilst for the $x_2$ component the force is
given by plus the gradient of the potential. Therefore, the
criteria for the stability of motion for the $x_2$ degree of
freedom is that the potential has to have a maximum in the
$(x_2,V)$ plane. This is just opposite to the case of the $x_1$
degree of freedom where stability requires a minimum for the
potential in the $(x_1,V)$ plane.
\section{Classical solutions}
\subsection{Commutative case}
Let us start by supposing that the dynamical variables defined in
(\ref{9}) and their conjugate momenta satisfy \cite{17}
\begin{equation}\label{15}
\{x_{\mu},p_{\nu}\}_{P}=\eta_{{\mu}{\nu}},
\end{equation}
where $\eta_{{\mu}{\nu}}$ is the two dimensional Minkowski metric
and $\{ , \}_{P}$ represents the Poisson bracket. In view of the
above, Hamiltonian (\ref{14}) can be written as
\begin{equation}\label{151}
{\cal H}=\frac{1}{4}\eta^{\mu\nu}p_{\mu}p_{\nu}+
\omega^2\eta^{\mu\nu}x_{\mu}x_{\nu},
\end{equation}
and equations of motion become
\begin{eqnarray}\label{16}
\left\{
\begin{array}{lll}
\dot{x_\mu}=\{x_\mu,{\cal
H}\}_{P}=\frac{1}{2}p_\mu, \\
\\
\dot{p_\mu}=\{p_\mu,{\cal H}\}_{P}=-2\omega^2x_\mu,
\end{array}
\right.
\end{eqnarray}
where in obtaining these equations we have used relations
(\ref{15}). Now, using equations (\ref{16})  we obtain
\begin{equation}\label{17}
\ddot{x_\mu}+\omega^{2}x_\mu=0.
\end{equation}
The solution of the above equations reads
\begin{equation}\label{18}
x_\mu(t)=A_\mu e^{i\omega t}+B_\mu e^{-i\omega t},
\end{equation}
where $A_\mu$ and $B_\mu$ are constants of integration. We note
that  Hamiltonian constraint (${\cal H}=0$) imposes the following
relation on these constants
\begin{equation}\label{19}
A_\mu B^\mu =0.
\end{equation}
Finally, using equations (\ref{9}) and (\ref{6}), the scale
factors take on the following forms for $d\neq3$
\begin{equation}\label{191}
\left\{
\begin{array}{lll}
a(t)=k_1 \left[\sin(\omega
t+\phi_1)\right]^{\frac{-2(\sigma_{+})\sqrt{\frac{d+2}{3d}}}{d-3}}
\left[\sin(\omega t+\phi_2)\right]^{\frac{2(\sigma_{-})\sqrt{\frac{d+2}{3d}}}{d-3}}, \\
\\
R(t)=k_2 \left[\sin(\omega
t+\phi_1)\right]^{\frac{2(\rho_{-})d\sqrt{\frac{d+2}{3d}}}{d-3}}\left[\sin(\omega
t+\phi_2)\right]^{\frac{-2(\rho_{+})d\sqrt{\frac{d+2}{3d}}}{d-3}},
\end{array}
\right.
\end{equation}
where $k_1$ and $k_2$  are arbitrary constants and $\phi_1$ and
$\phi_2$ are arbitrary phases. Note that if $\omega^2$ is
negative, the trigonometric functions are replaced by their
hyperbolic counterparts in the above solutions. The sign of
$\omega^2$ relates to the sign of the cosmological constant
(\ref{1a}). For $d=3$, similar solutions are easily found from
transformations (\ref{101}).
\subsection{Non-commutative case}
We now concentrate on the non-commutativity concepts with Moyal
product in phase space. The Moyal product in phase space may be
traced to an early intuition by Wigner \cite{wig} which has been
developing over the past decades \cite{gozzi}. Non-commutativity
in classical physics \cite{15} is described by the Moyal product
law between two arbitrary functions of position and momenta as
\begin{equation}\label{35}
(f\star_\alpha
g)(x)=\exp\left[\frac{1}{2}\alpha^{ab}\partial_a^{(1)}\partial_b^{(2)}\right]
f(x_1)g(x_2)|_{x_1=x_2=x},
\end{equation}
such that
\begin{eqnarray}\label{36}
\alpha_{ab}=\left(%
\begin{array}{cc}
 \theta_{\mu\nu}  & \eta_{\mu\nu}+\sigma_{\mu\nu} \\
  -\eta_{\mu\nu}-\sigma_{\mu\nu} & \beta_{\mu\nu}\\
\end{array}%
\right),
\end{eqnarray}
where the $N\times N$ matrices $\theta$ and $\beta$ are assumed to
be antisymmetric with $2N$ being the dimension of the classical
phase space and $\sigma$ can be written as a combination of
$\theta$ and $\beta$. With this product law, the deformed Poisson
brackets can be written as
\begin{equation}\label{37}
\{f,g\}_\alpha=f\star_\alpha g-g\star_\alpha f.
\end{equation}
A simple calculation shows that
\begin{equation}\label{38}
\begin{array}{lll}
\{x_\mu,x_\nu\}_\alpha=\theta_{\mu\nu}, \\
\{x_\mu,p_\nu\}_\alpha=\eta_{\mu\nu}+\sigma_{\mu\nu}, \\
\{p_\mu,p_\nu\}_\alpha=\beta_{\mu\nu} \hspace{.15cm}.
\end{array}
\end{equation}
It is worth noticing at this stage that in addition to
non-commutativity in $(x_1,x_2)$ we have also considered
non-commutativity in the corresponding momenta. This should be
interesting since its existence is in fact due essentially to the
existence of non-commutativity on the space sector \cite{gozzi,15}
and it would somehow be natural to include it in our
considerations.

Now, consider the following transformation on the classical phase
space $(x_\mu,p_\mu)$
\begin{equation}\label{39}
\begin{array}{ll}
{x_\mu'}=x_\mu-\frac{1}{2}\theta_{\mu\nu} p^\nu, \\
{p_\mu'}=p_\mu+\frac{1}{2}\beta_{\mu\nu} x^\nu
 .
\end{array}
\end{equation}
It can easily be checked that if $(x_\mu,p_\mu)$ obey the usual
Poisson algebra (\ref{15}), then
\begin{equation}\label{40}
\begin{array}{lll}
\{{x_\mu'},{x_\nu'}\}_{P}=\theta_{\mu\nu},\\
\{{x_\mu'},{p_\nu'}\}_{P}=\eta_{\mu\nu}+\sigma_{\mu\nu},\\
\{{p_\mu'},{p_\nu'}\}_{P}=\beta_{\mu\nu}.
\end{array}
\end{equation}
These commutation relations are the same as (\ref{38}).
Consequently, for introducing non-commutativity, it is more
convenient to work with Poisson brackets (\ref{40}) than
$\alpha$-star deformed Poisson brackets (\ref{38}). It is
important to note that the relations represented by equations
(\ref{38}) are defined in the spirit of the Moyal product given
above. However, in the relations defined in (\ref{40}), the
variables $(x_\mu,p_\mu)$ obey the usual Poisson bracket relations
so that the two sets of deformed and ordinary Poisson brackets
represented by relations (\ref{38}) and (\ref{40})  should be
considered as distinct.

\begin{figure}
\begin{center} \epsfig{figure=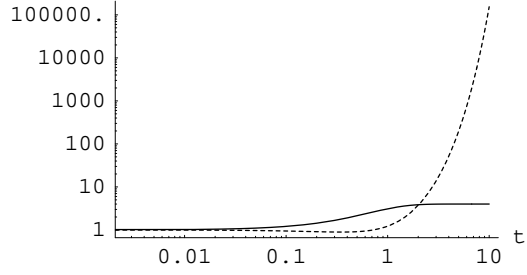,width=7cm}
\end{center}
\caption{\footnotesize Scale factor of the extra dimensions, solid
line and that of the ordinary space, dotted line. The values of
$\alpha$ and $\gamma$  are such that equation (\ref{71}) is
satisfied and $d=1$. } \label{fig1}
\end{figure}

Let us change the commutative Hamiltonian (\ref{151}) with minimal
variation to
\begin{equation}\label{41}
{\cal
H'}=\frac{1}{4}\eta^{\mu\nu}p'_{\mu}p'_{\nu}+\omega^2\eta^{\mu\nu}x'_{\mu}x'_{\nu},
\end{equation}
where we have the commutation relations
\begin{equation}\label{141}
\begin{array}{lll}
\{{x_\mu'},{x_\nu'}\}_{P}=\theta \epsilon_{\mu\nu},\\
\{{x_\mu'},{p'_{\nu}}\}_{P}=(1+\sigma)\eta_{\mu\nu},\\
\{{p_\mu'},p_\nu'\}_{P}=\beta \epsilon_{\mu\nu},
\end{array}
\end{equation}
with $\epsilon_{\mu\nu}$ being a totally anti-symmetric tensor and
$\sigma$ is given by
\begin{equation}\label{143}
\sigma=\frac{1}{4}\beta\theta.
\end{equation}
We have also set $\theta_{\mu\nu}=\theta \epsilon_{\mu\nu}$ and
$\beta_{\mu\nu}=\beta \epsilon_{\mu\nu}$. Using the transformation
introduced in (\ref{39}), Hamiltonian (\ref{41}) becomes
\begin{equation}\label{43}
{\cal
H}=\frac{1}{4}\left(1-\omega^2\theta^2\right)\eta^{\mu\nu}p_\mu
p_\nu+\left(\omega^2-\frac{\beta^{2}}{16}\right)\eta^{\mu\nu}x_\mu
x_\nu
-\left(\frac{\beta}{4}+\theta\omega^2\right)\epsilon^{\mu\nu}x_\mu
p_\nu.
\end{equation}
 The equations of motion
corresponding to  Hamiltonian (\ref{43}) are
\begin{equation}\label{44}
\left\{
\begin{array}{lll}
\dot{x_\mu}=\{x_\mu,{\cal
H}\}_{P}=\frac{1}{2}\left(1-\omega^2\theta^2\right)p_\mu+
\left(\frac{\beta}{4}+\theta\omega^2\right)\epsilon_{\mu\nu} x^\nu,\\
\\
\dot{p_\mu}=\{p_\mu,{\cal
H}\}_{P}=-2\left(\omega^2-\frac{\beta^{2}}{16}\right)x_\mu+
\left(\frac{\beta}{4}+\theta\omega^2\right) \epsilon_{\mu\nu} p^\nu,\\
\end{array}
\right.
\end{equation}
where we have used relations (\ref{15}). It can again be easily
checked that if one writes the equations of motion for
non-commutative variables, equations (\ref{141}), with respect to
Hamiltonian (\ref{41}) and  uses transformation rules (\ref{39}),
one gets a linear combination of the equations of motion
(\ref{44}). This points to the fact that these two approaches are
equivalent. Now, as a consequence of equations of motion
(\ref{44}), one has
\begin{equation}\label{45}
\ddot{x}_\mu-2\left(\frac{\beta}{4}+\theta\omega^2\right)\epsilon_{\mu\nu}\dot{x}^\nu
+\left[\left(1-\omega^2\theta^2\right)\left(\omega^2-\frac{\beta^2}{16}\right)+\left(\frac{\beta}{4}+
\theta\omega^2\right)^2\right]x_\mu=0.
\end{equation}
Note that upon setting $\theta=\beta=0$, we get back equation
(\ref{17}). Here, the equations have an additional velocity term
which may be interpreted as viscosity. The solution to equation
(\ref{45}) can be written as follows
\begin{equation}\label{46}
\left\{
\begin{array}{lll}
x_1(t)=Ae^{(-\alpha+i\gamma)t}+Be^{(-\alpha-i\gamma)t}+
Ce^{(\alpha+i\gamma)t}+De^{(\alpha-i\gamma)t}
, \\
\\
x_2(t)=Ae^{(-\alpha+i\gamma)t}+Be^{(-\alpha-i\gamma)t}-
Ce^{(\alpha+i\gamma)t}-De^{(\alpha-i\gamma)t} , \\
\end{array}
\right.
\end{equation}
where
\begin{equation}\label{47}
\alpha=\left(\theta\omega^2+\frac{\beta}{4}\right)\hspace{3mm}\mbox{and}\hspace{3mm}
\gamma=\left[\left(1-\omega^2\theta^2\right)\left(\omega^2-\frac{\beta^2}{16}\right)\right]^{1/2},
\end{equation}
with $A$, $B$, $C$  and $D$ being the constants of integration.
The Hamiltonian constraint, ${\cal H}=0$, leads to
\begin{equation}\label{48}
[\gamma+i\alpha]AD+[\gamma-i\alpha]BC=0.
\end{equation}
Setting $\theta=\beta=0$ in (\ref{48}) reproduces relation
(\ref{19}) in the commutative case, as one would expect. Also,
from equations (\ref{9}) and (\ref{6}) we recover the scale
factors as ($d\neq3$)
\begin{equation}\label{192}
\left\{
\begin{array}{lll}
a(t)=k_1 \left[\sin(\gamma
t+\phi_1)\right]^{\frac{-2(\sigma_+)\sqrt{\frac{d+2}{3d}}}{d-3}}\left[\sin(\gamma
t+\phi_2)\right]^{\frac{2(\sigma_-)\sqrt{\frac{d+2}{3d}}}{d-3}}
\exp\left[\frac{(6\alpha)\sqrt{\frac{d+2}{3d}}}{d-3}t\right], \\
\\
\ R(t)=k_2\left[\sin(\gamma
t+\phi_1)\right]^{\frac{2(\rho_-)d\sqrt{\frac{d+2}{3d}}}{d-3}}\left[\sin(\gamma
t+\phi_2)\right]^{\frac{-2(\rho_+)d\sqrt{\frac{d+2}{3d}}}{d-3}}\exp\left[\frac{-(2d\alpha
)\sqrt{\frac{d+2}{3d}}}{d-3}t\right],
\end{array}
\right.
\end{equation}
where, $k_1$ and $k_2$ are constants, $\alpha$ and $\gamma$ are
defined in (\ref{47}) and $\phi_1$ and $\phi_2$ are arbitrary
phases. Note that in the commutative case the exponential terms
were not present. Also, it is worth noticing that the first
equation in (\ref{192}) shows that if $\frac{\alpha}{d-3}$ is
negative then the scale factor corresponding to the extra
dimensions compactifies to zero as time increases. Again, as in
the  previous case, if $\gamma^2$ is negative, then in the above
solutions hyperbolic functions would replace the corresponding
trigonometric ones. In this case the above  scale factors for late
times become

\begin{equation}\label{193}
\left\{
\begin{array}{lll}
a(t)=k_1
e^{\frac{-2(\gamma\sigma_+)\sqrt{\frac{d+2}{3d}}}{d-3}t}
e^{\frac{2(\gamma\sigma_-)\sqrt{\frac{d+2}{3d}}}{d-3}t}
e^{\frac{(6\alpha)\sqrt{\frac{d+2}{3d}}}{d-3}t}, \\
\hspace{11cm}t\rightarrow+\infty
\\
R(t)=k_2e^{\frac{2(\gamma\rho_-)d\sqrt{\frac{d+2}{3d}}}{d-3}t}
e^{\frac{-2(\gamma\rho_+)d\sqrt{\frac{d+2}{3d}}}{d-3}t}
e^{\frac{-(2d\alpha)\sqrt{\frac{d+2}{3d}}}{d-3}t}.
\end{array}
\right.
\end{equation}
To make $a(t)$ stabilized, the total exponential in the first
equation should become finite. To this end we put the sum of the
exponents equal to zero. One thus finds a relation between
$\alpha$ and $\gamma$ as
\begin{equation}\label{71}
\alpha=\frac{\gamma}{3}(\sigma_+-\sigma_-).
\end{equation}
This equation can now be used for choosing (tuning) the scale
factors  such that $a(t)$ approaches a finite value for large $t$
while the scale factor representing the universe, $R(t)$, grows
exponentially. This points to the stabilization of extra
dimensions and can also be interpreted as their compactification
relative to the scale factor of the universe, see figure
\ref{fig1}. Note that for $d=3$, the relation (\ref{71}) also
holds.

At this point it would be appropriate to address the cosmological
constant problem, the huge disparity between the values of the
cosmological constant predicted in cosmology and particle physics.
A number of approaches have been suggested to address this problem
\cite{16}. However, as will be shown below, this problem has a
natural solution in the non-commutative approach. We note from
above that $\omega^2$ is proportional to the cosmological constant
$\Lambda$. Comparing equations (\ref{1a}), (\ref{191}), (\ref{47})
and (\ref{192}) we propose that the relation (\ref{1a}) between
the oscillating frequency $\gamma$ and the cosmological constant
$\Lambda_{nc}$ continues to hold, that is
\begin{equation}\label{66}
\Lambda_{nc}=2\left(\frac{d+2}{d+3}\right)\gamma^2=
2\left(\frac{d+2}{d+3}\right)\left(1-\omega^2\theta^2\right)
\left(\omega^2-\frac{\beta^2}{16}\right).
\end{equation}
As has already been noticed, the non-commutativity concept in our
model is closely related to extra dimensions and will disappear if
the extra dimensions  disappear, that is if we set $d=0$. Since
the appearance of extra dimensional effects could be more
naturally attributed to short distance scales, it would be
reasonable to assume that $\Lambda_{nc}$ is the cosmological
constant associated with particle physics. Now, from equations
(\ref{66}) and (\ref{1a}) one has
\begin{equation}\label{67}
\Lambda_{nc}=\left[1-\frac{1}{2}\left(\frac{d+3}{d+2}\right)\theta^2\Lambda\right]
\left[\Lambda-\frac{1}{8}\left(\frac{d+2}{d+3}\right)\beta^2\right].
\end{equation}
Since the cosmological constant $\Lambda$ appearing in the above
equation is that representing large scales and has consequently a
very small value, we see from equation (\ref{67}) that the
particle physics cosmological constant is
\begin{equation}\label{68}
\Lambda_{nc}\sim-\frac{1}{8}\left(\frac{d+2}{d+3}\right)\beta^2.
\end{equation}
If $\beta^2$ is large relative to $\Lambda$ then from (\ref{68})
$\Lambda_{nc}$ will become much larger than $\Lambda$. The above
analysis points  to a possible solution of the long standing
cosmological constant problem. The above argument is also true up
to a multiplicative constant for the case $d=3$.

To have a geometrical understanding of non-commutativity of the
scale factors, let us see how such a relation looks. One may show
that for $d\neq3$
\begin{eqnarray}
\{\Phi,R\}_\alpha=\frac{4d\left(\frac{d+2}{3d}\right)^{1/2}}{d-3}\frac{\theta}{R^2},
\end{eqnarray}
while for $d=3$
\begin{eqnarray}
\{\Phi,R\}_\alpha=-\frac{5\sqrt{5}}{216}\frac{\theta}{R^2}.
\end{eqnarray}
These relations show that, no matter how large $\theta$ may get,
the effect of non-commutativity would be very small in the present
epoch due to the large value of the scale factor $R$.
\section{Conclusions}
In this paper we have introduced non-commutativity between scale
factors of the ordinary universe and extra dimensions in a
multi-dimensional cosmological model. We have shown that the
classical solutions of such a model clearly point to a possible
resolution of the cosmological constant problem and
compactification of extra dimensions. We have also shown  that in
this model the two parameters representing non-commutativity are
not independent and satisfy a certain relationship, equation
(\ref{71}), which may be used for the purpose of fine-tuning so as
to make the extra dimensions stabilized.
\vspace{10mm}\noindent\\
{\bf Acknowledgement}\vspace{2mm}\noindent\\
The authors would like to thank the anonymous referee for valuable
comments.

\end{document}